\documentclass[reprint,showpacs,amsfonts,amsmath,amssymb,superscriptaddress,prb]{revtex4-1}
\usepackage{graphicx}
\usepackage{dcolumn}
\usepackage{bm}
\usepackage{color}

\begin{document}

\title{Projective symmetry group classification of $Z_3$ parafermion spin liquids on a honeycomb lattice}

\author{Zhao-Yang Dong}
\affiliation{National Laboratory of Solid State Microstructures and Department of Physics, Nanjing University, Nanjing 210093, China}
\author{Shun-Li Yu}
\affiliation{National Laboratory of Solid State Microstructures and Department of Physics, Nanjing University, Nanjing 210093, China}
\affiliation{Collaborative Innovation Center of Advanced Microstructures, Nanjing University, Nanjing 210093, China}
\author{Jian-Xin Li}
\email{jxli@nju.edu.cn}
\affiliation{National Laboratory of Solid State Microstructures and Department of Physics, Nanjing University, Nanjing 210093, China}
\affiliation{Collaborative Innovation Center of Advanced Microstructures, Nanjing University, Nanjing 210093, China}

\date{\today}

\begin{abstract}
To study exotic excitations described by parafermions in the possible spin liquid states of SU($n$) spin systems, we introduce a parafermion parton approach. The SU($n$) spin operators can be represented by clock and shift matrices, which are shown to be the polynomials of parafermion operators in the parafermion representation. We find that SU($n$) spins can be decomposed into $n$ parafermion matrices of degree one. In this decomposition, the spin has a $\{\bigotimes{\rm SU}(n)\}^{n-1}$ gauge symmetry. As an application, we study the one-dimensional three-state clock model and generalized Kitaev model by a mean-field theory, both of them have been proved to be related to parafermion excitations. We find that with the symmetries of translations, $6$-fold rotation and combination of parity and time reversal, there are $9$ types and $102$ solutions for two-dimensional $Z_3$ parafermion spin liquids on the honeycomb lattice. On the contrast, there are $9$ types and $36$ solutions if both parity and time-reversal symmetries are present. Our results provide a novel route for the systematic search of new types of spin liquids with exotic anyon excitations.
\end{abstract}


\maketitle

\section{INTRODUCTION}
In the beginning of this century, Kitaev gave an elegant way of understanding the phases and physical implications of the $1$-dimensional(D) transverse-field Ising model in the fermionized version \cite{Kitaev2001}. In the fermionic representation, the quantum phase transition in this model can be understood as a transition from the weak pairing BCS regime to the strong pairing BEC regime. The weak-pairing phase is topologically non-trivial and a chain with open boundaries possesses a Majorana zero-energy mode localized at each end.
Not long after that, Kitaev proposed the so-called \emph{Kitaev honeycomb model} (or simply the \emph{Kitaev model})\cite{Kitaev2006}, which is one of the famous examples
of exactly solvable \emph{spin} models in theoretical condensed matter physics. This model consists of $s=1/2$ spins, and their interactions between the nearest neighbors are of $xx$-, $yy$- or $zz$-type Ising couplings, depending on the directions of links. Its exact ground state has a gapless or gapped spin liquid phase, depending on the interaction parameters, respectively. The elementary excitations are described by itinerant Majorana fermions coupled with a $Z_2$ gauge field. Kitaev had showed that the gapless phase acquires a gap in the presence of magnetic field and the excitations are non-Abelian anyons.

Kitaev's insight inspires many similar spin models described by Majorana fermions with non-Abelian anyon excitations soon afterwards\cite{Yao2007,Yao2009,Wu2009}. Furthermore, the generalizations of the Kitaev model to a much larger class of partially integrable spin models have also been introduced\cite{Vaezi2014,Barkeshli2015}. The resulting Hamiltonians have many interesting properties similar to Kitaev's original model and could be related to the $2$-D parafermion systems coupled to a discrete gauge symmetry. Parafermions, the generalization of Majorana fermions\cite{Fendley2012,Mong2014,Fendley2014,Alicea2016}, attract considerable attention recently owing to their potential utility for a universal and intrinsically topological quantum computation\cite{Nayak2008,Alicea2016} based on their non-Abelian statistics with a $2\pi/n$ phase. Parafermion commutation relations have appeared in literatures long ago\cite{120000874548,MORRIS1967,Fradkin1980}. The concept of parafermions have also been used in the conformal field theory to describe critical points in $Z_n$-symmetrical statistical systems\cite{Zamolodchikov_Fateev_1985,Mong2014}. Moreover, the potential realizations of parafermion zero modes are also predicted in the Read-Rezayi state for the fractional quantum Hall effect (FQHE)\cite{Read1999,Xia2004,Rezayi2009}, including edge modes in FQHE\cite{Lindner2012,Clarke2013,Barkeshli2014} and fractional topological insulators\cite{Cheng2012,Vaezi2013,Burrello2013}.
Based on these progress in the studies of spin liquids and parafermions, it is natural to ask whether we might find spin-liquid states with non-trivial topological orders and parafermion excitations.

As mentioned above, Majorana fermions are elementary excitations of spin-liquid states in the Kitaev model with $s=1/2$ spins, so we can expect that parafermions might be the fractionalized quasi-particles of a Kitaev-like model with large spins ($s>1/2$). However, a large spin will suppress the quantum fluctuations which are essential to melt the magnetic moment and realize the spin-liquid state.
An alternative route to preserve quantum fluctuations is to extend the SU($2$) spin to SU($n$). As $n$ increases from $2$ to larger values, quantum spin fluctuations are enhanced\cite{Misguich2003,Hermele2009,Hermele2011}. Such fluctuations may melt any form of classical orders and lead to various exotic spin-liquid states.
Therefore, exotic excitations that could be described by parafermions are expected to exist in the spin-liquid states of SU($n$) spin systems.
Moreover, we note that the studies of spin-liquid states of SU($n$) spins are not only purely theoretical exercises but also relevant to experimental realizations, as the SU($n$) spins can be realized in cold atom systems\cite{Gorshkov2010,Pagano2014,Scazza2014,XZhang2014,Cazalilla2014}, in quantum dot arrays\cite{Onufriev1999}, or in materials with strong spin-orbital couplings\cite{KarloPenc2003}.

A microscopic theory of spin-liquid states should involve fractionalized excitations, so the mean-field theory is a natural and simplest approach for an initial study, which can provide qualitative results and insights on the possible spin-liquid states. 
Although the mean-field Hamiltonian usually partially or completely breaks the symmetries of the spin Hamiltonian, two apparently
different mean-field solutions differing by a gauge transformation may describe the same physical state after projected onto the physical space.
Therefore, a classification method for the mean-field spin-liquid states is needed. The projective symmetry group (PSG) classification for spin-liquid states with fermionic fractionalization on the square lattice was introduced by Wen ~\cite{Wen2002}. 
The PSG classification seeks to list all the possible classes of lattice symmetry representations in the enlarged Hilbert space of fractionalized spinons, which is an enumeration and characterization of possible quantum spin liquid phases. The extensions to other spin models with different symmetries\cite{Lu2011a,Lu2011,PhysRevB.92.060407,Bieri2016,PhysRevLett.118.267201}, and other representations based on Majorana fermions\cite{You2012,Chen2012} and bosonic spinons\cite{Wang2006,Wang2010,PhysRevB.87.125127} have also be made.

The main purpose of our paper is to present a systematic PSG classification for the SU($n$) spin liquid using the fractionalization with parafermions. We first introduce a general parafermion parton approach to the SU($n$) spins. We find that a SU($n$) spin can be interpreted as a confinement of $n$ parafermions in the parafermion representation and the Hilbert space is extended from $n$ dimensions of a local SU($n$) spin to $n^2$ dimensions of four parafermion operators. It is easy to verify that there are $n-1$ SU($n$) gauge symmetries, i.e. there are $n-1$ gauge transformations, under which spin operators are invariant. We then present the projective realizations of possible symmetries of the SU($n$) spin lattice models based on the mean field theory with parafermion parton approach. As an example, $Z_3$ PSGs on the honeycomb lattice are calculated. There are $9$ types and $102$ solutions with the symmetries of translations, $6$-fold rotation and combination of parity and time reversal. Furthermore, there are $9$ types and $36$ solutions if both parity and time-reversal symmetries are present.

The paper is organized in the following manner. In section \ref{pre}, we briefly review some notations and properties of the parafermion operators, clock and shift (C\&S) matrices, which are mathematical basis of this paper. In section \ref{parton}, the general parafermion parton approach to SU($n$) spin operators is introduced. In section \ref{liquid}, we provide a mean-field theory in the parafermion representation for parafermion spin liquids, and discuss the constraint they impose on the mean-field Hamiltonian. Then, two well-known models, the clock model and generalized Kitaev model, are studied by the mean-field theory. In section \ref{PSG}, we present the general theory of parafermion projective symmetry representations, and construct the possible PSGs of the generalized Kitaev model as an example. Finally, in section \ref{Z3}, we present all possibilities of the $Z_3$ PSGs on the honeycomb lattice, and the detailed calculations are in the appendix.

\section{BASIC CONCEPTS}\label{pre}

Before getting into the main subject of this paper, we are going to simply review the basic properties and some common conceptions of parafermions. Then, we will introduce the C\&S matrices and their application to SU($n$) spins.

\subsection{Parafermions}

Parafermions was defined from string operators of Z($N$) spins in the Z($N$) theory long ago\cite{PhysRevD.21.2878,PhysRevD.24.1562}.
Because of potential relevance to topological quantum information processing, theoretical studies about parafermions still made progress recently\cite{Cobanera2014,Jaffe2015}.
Parafermions can be regarded as the simplest generalization of Majorana fermions. A well-known characteristic of the Majorana fermion is that it is its own antiparticle, so the creation and annihilation operators of Majorana fermions are identical, i.e., $\gamma^\dag=\gamma$. Moreover, the Majorana operators satisfy the following Clifford algebra:
\begin{equation}
\gamma_i^2=1,\quad \gamma_i \gamma_j=-\gamma_j \gamma_i.
\end{equation}
We can generalize the above algebra of Majorana fermions to the generalized Clifford algebra (GCA)\cite{doi:10.1063/1.526853,jagannathan2010generalized} of order $n$,
\begin{equation}
  \gamma_i^n=1,\quad\gamma_i \gamma_j=\omega^{\mathfrak{sgn}(j-i)}\gamma_j \gamma_i,
  \label{eq1}
\end{equation}
where $\mathfrak{sgn}$ is the sign function and $\omega=e^{2\pi i/n}$. In this case, $\gamma^\dag=\gamma^{-1}=\gamma^{n-1}$, which means that the creation operator $\gamma^{\dag}$ and annihilation operator $\gamma$ are different unless $n=2$ in which they are reduced to a self-adjoint representation of a Clifford algebra as in the Majorana case. For $n\geq3$, $\gamma_{i}$'s described by Eqs. (\ref{eq1}) are called the \textit{parafermion generators of order n} or simply the \textit{parafermion operators}.
For a finite system, the number $L$ of parafermion operators are even, and the operators have an unique irreducible representation on a Hilbert space $\mathcal{H}$ of dimension $N = n^{L/2}$. For Majorana operators, this is obvious through expressing them in terms of fermionic annihilation and creation operators.

The ordered monomials of the parafermion operators are expressed as
\begin{equation}
  m_\mathfrak{I}^{[i]}=\gamma_1^{n_1}\gamma_2^{n_2} \cdots \gamma_L^{n_L},
  \label{eq2}
\end{equation}
where $0\leq n_j\leq n-1$, and they can be denoted by the sets of integers $\mathfrak{I} =\{n_1, n_2, \cdots,n_L\}$. The degree of the monomial is defined as
\begin{equation}
  i=\left(\sum\nolimits_j n_j\right)~\mathbf{mod}~n.
  \label{eq3}
\end{equation}
When the polynomials are composed of the monomials with the same degree $[i]$, we use $p^{[i]}$ to indicate the degree of the polynomials.

The aforementioned points are the basic properties of parafermion operators that we will use in the paper. 

\subsection{C\&S matrices and their application to SU($n$) spin}

The clock and shift matrices are first introduced by Sylvester\cite{Sylvester1882,*Sylvester1883,*Sylvester1884}. Their famous utilization is in the $n$-state Potts model\cite{WuFY1982}, which is a generalization of the Ising model by replacing a two-state Ising spin with a ``spin" of $n$ states. The basic operators of the clock and shift matrices, denoted by $\sigma$ and $\tau$ respectively, generalize the Pauli matrices $\sigma_z$ and $\sigma_x$ to $n\times n$ matrices:
\begin{equation}
  \sigma=\left(
  \begin{array}{cccc}
    1 &   0     &   \cdots     & 0 \\
    0  & \omega &   \cdots     & 0  \\
    \vdots  & \vdots       & \ddots & \vdots  \\
    0  &     0   &   \ldots     & \omega^{n-1}
  \end{array}
  \right),\quad
  \tau=\left(
  \begin{array}{cccc}
    0  &  \ldots      &  0 & 1 \\
    1 &   \cdots     & 0  &  0 \\
    \vdots  & \ddots & \vdots  & \vdots  \\
    0 &  \ldots      & 1 & 0
  \end{array}
  \right),
  \label{eq4}
\end{equation}
where $\omega=e^{2\pi i/n}$. Similar to parafermion operators, $\sigma$ and $\tau$ fulfil the following GCA:
\begin{equation}
  \sigma^n=1,\quad\tau^n=1,\quad\sigma\tau=\omega\tau\sigma.
  \label{eq5}
\end{equation}
Besides, they also satisfy that $\sigma^{\dag}=\sigma^{-1}$ and $\tau^{\dag}=\tau^{-1}$. However, they are defined on the local $n$ states and commute with each other on different sites, which is consistent with ``spin" operators.

In order to study the SU($n$) spin system, we generalize usual SU($2$) spins to SU($n$) spins by the following C\&S matrices:
\begin{equation}
\begin{array}{cccc}
  1,            & \tau,             & \cdots, & \tau^{n-1}, \\
  \sigma,       & \sigma\tau,       & \cdots, & \sigma\tau^{n-1}, \\
  \vdots,       & \vdots,           & \ddots, & \vdots, \\
  \sigma^{n-1}, & \sigma^{n-1}\tau, & \cdots, & \sigma^{n-1}\tau^{n-1}.
\end{array}
  \label{eq6}
\end{equation}
These $n^2$ operators are obviously linearly independent. Thus, considering the space of local operators as a linear space of dimension $n^2$, any local operator can be expanded in the basis of these $n^2$ operators.

In fact, any SU($n$) transformation can also be formally expressed as a linear combination of $n^2$ linearly independent unitary matrices $\sigma^{a}\tau^{b}$, i.e.,
\begin{equation}
U=\sum_{a,b=1}^{n}c_{ab}\sigma^a\tau^b,\label{eq71}
\end{equation}
where $c_{ab}={\rm tr}(\tau^{-b}\sigma^{-a}U)/n$.
Thus $\sigma^a\tau^b$ can operate rotations of an SU($n$) spin. We also note that any element of SU($n$) can be expressed as an exponential of $n^2-1$ infinitesimal generators, which are traceless Hermitian matrices. So, one can construct $n^2-1$ generators from Eq.~(\ref{eq6}),
\begin{equation}
   \sigma^a\tau^b+\tau^{n-b}\sigma^{n-a},~ i(\sigma^a\tau^b-\tau^{n-b}\sigma^{n-a}),
   \label{eq7}
\end{equation}
whose commutation relations can be derived from $\sigma^a\tau^b\sigma^{a'}\tau^{b'}=\omega^{-a'b}\sigma^{a+a'}\tau^{b+b'}$.


\section{paraferMION PARTON APPROACH TO SU($n$) SPIN OPERATORS}\label{parton}

\subsection{Construction of the parafermion parton approach }

At the heart of a spin liquid construction is the fractionalization
of spin excitations in term of partons, i.e., effective low-energy excitations carrying a fractional spin quantum number.
For example, the celebrated Kitaev model is solved by writing the spin-1/2 operators in a Majorana fermion representation\cite{Kitaev2006}, which leads to an exact mean-field description for the spin liquid phase.
Recently, the parafermion representation for $Z_3$ rotors in the $Z_3$ generalization of the Kitaev model on triangular lattice\cite{PhysRevB.90.075106} and
the slave genons representation related to GCA for generalized Kitaev models\cite{Barkeshli2015} have been proposed. These parton representations succeed in describing the non-Abelian topological phase.

Here, we develop the parafermion parton approach for SU($n$) spins. In Sec.II B, we have generalized the usual SU($2$) spins to SU($n$) spins by the C\&S matrices given in Eq.(\ref{eq6}). So, our construction is carried out by representing the C\&S matrices by parafermions. We find that this representation has the following forms,
\begin{equation}
  \sigma_0={1\over n}\sum_{i=0}^{n-1}\sigma_{all}^{i+1}\sigma_{aid}^{n-i},
  ~\tau_0={1\over n}\sum_{i=0}^{n-1}\tau_{all}^{i+1}\tau_{aid}^{n-i}.
  \label{eq10}
\end{equation}
Here, two families of monomial operators are introduced to facilitate the parafermion representation: $\sigma_{all}=\omega^{(n-1)/2}\beta\gamma^{n-1},\tau_{all}=\omega^{(n-1)/2}\alpha^{n-1}\beta$ and $\sigma_{aid}=\omega^{(n-1)/2}s^{n-1}\alpha,\tau_{aid}=\omega^{(n-1)/2}s^{n-1}\gamma$.
$\alpha,\beta,\gamma,s$ are four parafermion operators of order $n$ with the local commutation relations,
\begin{eqnarray}
\alpha\beta = \omega\beta\alpha,~&& \beta\gamma = \omega\gamma\beta,\nonumber\\
\alpha\gamma = \omega\gamma\alpha,~&& \zeta s = \omega s\zeta,
\label{eq101}
\end{eqnarray}
where $\zeta=\alpha,\beta,\gamma$. According to Eq.~(\ref{eq1}), these parafermion operators can be written in order as $\alpha,\beta,\gamma,s$ (or $s,\alpha,\beta,\gamma$, if the commutation relation between $\zeta$ and $s$ is $\zeta s = \omega^{-1} s\zeta$). With Eq.(\ref{eq101}), one can easily find that the monomial operators
satisfy the GCA as expressed by Eq.~(\ref{eq5}), and the commutation relations between the two families,
\begin{eqnarray}
  &[\sigma_{all},\sigma_{aid}]=[\tau_{all},\tau_{aid}]=0,~\sigma_{all}\tau_{aid}=\sigma_{aid}\tau_{all},
  \nonumber\\
  &\sigma_{all}\tau_{aid}=\omega\tau_{aid}\sigma_{all},~\sigma_{aid}\tau_{all}=\omega\tau_{all}\sigma_{aid}.
  \label{eq9}
\end{eqnarray}

With the parafermion representation, the Hilbert space of dimension $n$ for a local SU($n$) spin has been enlarged to that of dimension $n^2$ of the four flavor parafermions. Noticing that the representations of $\sigma_{0}$ and $\tau_{0}$ in Eq.~(\ref{eq10}) in the Hilbert space of dimension $n^2$ are reducible with a fundamental $n$-D representation and $n^2-n$ trivial representations, we can impose the following local constraint to remove the trivial representations.
\begin{equation}
  \alpha^{n-1}\beta\gamma^{n-1}s=1,
  \label{eq11}
\end{equation}
It is equal to $\sigma_{all}=\sigma_{aid}$ (or $\tau_{all}=\tau_{aid}$).

Up to now, the formal construction of the parafermion parton approach is completed. 
To proceed, we would discuss the gauge symmetry in the redundant space of the representation, and the relation between the SU($n$) spin and four flavor parafermions.
For the SU($n$) spin, one can define $n-1$ SU($n$) transformations in the redundant $(n^2-n)$-D space under which Eq.~(\ref{eq10}) is invariant. Thus, they are related to the gauge symmetries, whose C\&S matrices are given by,
\begin{equation}
  \sigma_k={1\over n}\sum_{i=0}^{n-1}\omega^{ik}\sigma_{all}^{i+1}\sigma_{aid}^{n-i},
  ~\tau_k={1\over n}\sum_{i=0}^{n-1}\omega^{ik}\tau_{all}^{i+1}\tau_{aid}^{n-i},
  \label{eq12}
\end{equation}
for $k=1,2,\cdots,n-1$, and satisfy the commutation relations,
\begin{equation}
  \sigma_k\tau_k=\omega\tau_k\sigma_k,
  ~\sigma_k\sigma_l=\tau_k\tau_l=\sigma_k\tau_l=0,~{\rm for}~k\neq l.
  \label{eq13}
\end{equation} When $k=0$, they are C\&S matrices of the SU($n$) spin. The constraint Eq.~(\ref{eq11}) implies $\sigma_k=\tau_k=0$ for $k\neq0$.
With $\sigma_k$ and $\tau_k$, the Hilbert space of dimension $n^2$ can be divided into $n$ SU($n$) spaces which are generated by Eq.~(\ref{eq12}). Moreover, the summation $\sigma_{all}=\sum\sigma_k~(\tau_{all}=\sum\tau_k)$ generates a direct sum of $n$ SU($n$) spaces.
We also note that the C\&S matrices in Eq.~(\ref{eq12}) are polynomials of degree $[0]$ (also degree $[n]$), which ensures that the matrices on different sites are commutative, so they can be viewed as the SU($n$) spin operators.

On the other hand, from the view of four flavor parafermions, there will be $n$ SU($n$) transfermations generated by $\sigma_k,\tau_k$ and a combined SU($n$) transfermation generated by $\sigma_{all},\tau_{all}$. When restricted by the constraint Eq.~(\ref{eq11}), then an SU($n$) spin emerges.

Before concluding the construction, we would like to introduce parafermion matrices of degree $[1]$ to factorize Eq.~(\ref{eq10}), in order to show a clearer fractionalization representation of SU($n$) spins. It follows the similar procedure to represent a spin operator of $s=1/2$ by a bilinear, spinor, or even matrix (tensor) of partons (bosons or fermions).
In this way, an SU($n$) spin is a confinement of $n$ parafermions. According to Eq.~(\ref{eq71}), the generators of all $n$ SU($n$) transfermations can be expanded by $p_k^{[0]ab}=\sigma_k^a\tau_k^b$, and we define
\begin{equation}
  p_k^{[i]ab}=p_k^{[0]ab}s^i.
  \label{eq14}
\end{equation}
Consequently,
\begin{equation}
  p_{k+1}^{[i]ab}=s^{-1}p_k^{[i]ab}s,
  ~(p_k^{[i]ab})^\dagger=\omega^{-ab}p_{k+i}^{[-i]-a-b}.
  \label{eq15}
\end{equation}
Now the parafermion matrices are written as
\begin{equation}
  P_k^{[i]}=\sum_{a,b}\tau^a\sigma^bp_k^{[i]ab},
  \label{eq16}
\end{equation}
where $\sigma$ and $\tau$ are the clock and shift matrices as those in Eqs.~(\ref{eq4}). They satisfy $P_k^{[i]}P_l^{[j]}=\delta_{i+k,l}P_k^{[i+j]}$ and $(P_k^{[i]})^\dagger=P_{k+i}^{[-i]}$. Considering Eq.~(\ref{eq14}), (\ref{eq15}) and (\ref{eq16}), we can write the SU($n$) operators by the parafermion matrices,
\begin{equation}
  p_k^{[0]ab}=\sigma_k^a\tau_k^b={1\over n^{2(n-1)}}{\rm Tr}\left(\sigma^{-b}\tau^{-a}\prod_{i=0}^{n-1}P_{k+i}^{[1]}\right).
  \label{eq17}
\end{equation}
For any SU($n$) transformation generated by $p_k^{[0]ab}$, there is a matrix $U_k\in\rm{SU}(n)$: $P_k^{[i]}\rightarrow U_k^\dagger P_k^{[i]}$. From  Eq.~(\ref{eq17}), $p_0^{[0]ab}$ rotates as a vector under $U_0$, but is invariant under other transformations. Thus $U_0$ is the rotation of the SU($n$) spin, and the others are gauge transformations. It is the fact that the full local redundancy of the parafermion representation of an SU($n$) spin is the direct product of $n-1$ SU($n$).

\subsection{Parton approach for $s=1/2$ spin: a simplest example}

In order to better understand the parafermion parton approach, we next show how to carry out the Majorana representation for the $s=1/2$ spin using the method described above. The Schwinger-fermion representation of the $s=1/2$ spin can be written in a compact form as\cite{IanAffleck}
\begin{equation}
  \bm{S}=-{1\over 4}{\rm Tr}(\bm{\sigma}FF^\dagger),
  \label{eq18}
\end{equation}
where $\bm{\sigma}$ are the Pauli matrices and
\begin{equation}
  F=\left(
  \begin{array}{cc}
    f_\uparrow & f_\downarrow^\dagger \\
    f_\downarrow & -f_\uparrow^\dagger
  \end{array}
  \right).
  \label{eq19}
\end{equation}
Here, $f_{\alpha}$ ($f_\alpha^\dagger$) is an ordinary spin-$1/2$ fermionic annihilation (creatation) operator with spin $\alpha=\uparrow,\downarrow$. The convenience of this notation is that the left SU(2) rotations of $F$ are spin rotations, while the right SU(2) rotations are gauge transformations, under which Eq.~(\ref{eq18}) is totally invariant. Moreover, the gauge transformations are generated by
\begin{equation}
  \bm{G}={1\over 4}{\rm Tr}(F\bm{\sigma}F^\dagger).
  \label{eq20}
\end{equation}
In the representation of spin-1/2 fermions, the spin rotations are the same as those in the single-particle states but trivial in the vacuum and doubly-occupied states (which is the spin singlet states), while the gauge transformations are SU($2$) transformations in the manifold of particles. It seems that the relationship between them is almost the same as that in the parafermion representation. Obviously, they can also be combined to form a combination SU($2$) transformation $\bm{T}=\bm{S}+\bm{G}$.

The Majorana operators can be defined as
\begin{eqnarray}
  \alpha &=& f_\downarrow+f_\downarrow^\dagger, \nonumber\\
  \beta &=& -i(f_\downarrow-f_\downarrow^\dagger), \nonumber\\
  \gamma &=& f_\uparrow+f_\uparrow^\dagger, \nonumber\\
  s &=& -i(f_\uparrow-f_\uparrow^\dagger).
  \label{eq21}
\end{eqnarray}
Consequently,
\begin{equation}
  F={1\over 2}(is+\alpha\sigma^x+\beta\sigma^y+\gamma\sigma^z).
  \label{eq22}
\end{equation}
Comparing with Eq.~(\ref{eq16}), we find that $F$ is the parafermion matrix. Thus, Eq.~(\ref{eq18}) and (\ref{eq20}) are nothing but a specific form of Eq.~(\ref{eq17}) in the SU($2$) case. Written in the form of Eq.~(\ref{eq12}), Eq.~(\ref{eq18}) and (\ref{eq20}) read
\begin{eqnarray}
  S^\sigma &=& -{i\over 4}\left(s\sigma+\alpha\beta\gamma\sigma\right), \label{eq23}\\
  G^\sigma &=& {i\over 4}\left(s\sigma-\alpha\beta\gamma\sigma\right), \label{eq24}
\end{eqnarray}
where the superscripts $\sigma=x,y,z$ and the operators $\sigma=\alpha,\beta,\gamma$. Correspondingly, $\sigma_{all}=2T^\alpha=i\beta\gamma,\sigma_{aid}=2(S^\alpha-G^\alpha)=is\alpha$, $\tau_{all}=2T^\gamma=i\alpha\beta$ and $\tau_{aid}=2(S^\gamma-G^\gamma)=is\gamma$. The constraint $\alpha\beta\gamma s=1$ for the Majorana representation of the $s=1/2$ spin also comply with the form of Eq.~(\ref{eq11}). In a word, all of the properties are consistent with those generally stated in the parafermion parton approach.

Before ending this section, we emphasize that we can certainly write an SU($n$) spin in the fermion representation,
\begin{equation*}
  S^{\alpha\beta}=c^\dagger_\alpha c_\beta,
\end{equation*}
where $c^\dagger_\alpha$ ($c_\beta$) are fermionic creation (annihilation) operators, and $S^{\alpha\beta}$ are the SU($n$) generators which satisfy $[S^{\alpha\beta}, S^{\alpha'\beta'}]=\delta_{\beta\alpha'} S^{\alpha\beta'}-\delta_{\beta'\alpha}S^{\alpha'\beta}$.
In this representation, the dimension of space is enlarged to $2^{n}$, the redundancy is exponential growth and the particle-hole symmetry will suffer in the cases of $s>1/2$\cite{Liu2010}.

\section{Mean-field Theory of Parafermion SPIN LIQUIDS}\label{liquid}

With a complete parafermion representation theory, we are well-prepared for a mean-field theory.
Although the mean-field theory can hardly give the quantitative information about the low-energy properties of a system, it does qualitatively tell us the basic properties of spin-liquid states\cite{Wen2002}. In the mean-field theory, if we rewrite the SU($n$) spin operators in the form of parafermion partons and construct the low-energy effective model, the emergent parafermion excitations are expected. The physical spin-liquid states can be obtained by projecting the mean-field states to the physical subspace. Moreover, with the help of PSG, such spin liquid states can be classified\cite{Wen2002}.

In this paper, we concentrate on the SU($n$) spin models on the 1-D or 2-D lattices,
\begin{equation}
  H=\sum_{r,r'}J^{rr'}_{aba'b'}p_{0r}^{[0]ab}p_{0r'}^{[0]a'b'}+\cdots+h.c.,
  \label{eq41}
\end{equation}
where $p_{0}^{[0]ab}$ is the generators of an SU($n$) group defined before, $r$ and $r'$ are the lattice indices, and the repeated spin indices $a$, $b$, $a'$ and $b'$ are summed over the two spin operators.
Since we mainly focus on the approach and generic properties of the SU($n$) spin liquids rather than a concrete model, here we do not consider the complex interactions involving more than two spin operators (such as the ring-exchange interactions).
Since the spin operators are written as the polynomials of degree $[0]$, every term in Eq.~(\ref{eq41}) is degree $[0]$. In the mean-field theory, we decompose $p_{0r}^{[0]ab}p_{0r'}^{[0]a'b'}$ into two polynomials of degree $[0]$: $p_{0r}^{[0]ab}p_{0r'}^{[0]a'b'}=AB$ ($A$ and $B$ are commutative, i.e., $[A,B]=0$), and we introduce bosonic fields to decouple such two-polynomial interactions. In the saddle-point approximation, the bosonic fields are related to the correlations: $AB=\chi_B A+\chi_A B-\chi_A\chi_B$, where $\chi_A=\langle A\rangle$ and $\chi_B=\langle B\rangle$. Therefore, the mean-field Hamiltonian for Eq.~(\ref{eq41}) can be written as
\begin{equation}
  H_0=\sum_{r,r'}\chi^{rr'}_{\zeta\xi}\zeta_r^m\xi_{r'}^{n-m}+h.c.\cdots,
  \label{eq44}
\end{equation}
where $\zeta,\xi=\alpha,\beta,\gamma,s$.

To perform the mean-field theory, we have to enlarge the Hilbert space. In this way, the mean-field Hamiltonian can hardly give the true ground state energy, even the mean-field ground state is not a valid wave function for the spin system if there is no constraint.  
In this case, in order to obtain the valid state, we need to project the mean-field wave function to the space with the constraint Eq.~(\ref{eq11}),
\begin{equation}
  \Psi_{spin}=P\Psi_{mean},~P=\sum_{i=0}^{n-1}(\alpha^{n-1}\beta\gamma^{n-1}s)^i.
  \label{eq42}
\end{equation}
The projector can be obtained by $P=(p_{0r}^{[0]ab})^n$. The projector and gauge transformations are commutative, and the gauge transformations has no effect on $\Psi_{spin}$, so
\begin{equation}
  \Psi_{spin}=U_g\Psi_{spin}=P\Psi_{mean}=U_gP\Psi_{mean}=PU_g\Psi_{mean}.
  \label{eq43}
\end{equation}
It means that different mean-field wave functions related by a gauge transformation give rise to the same physical spin state after the projection. In addition, we can introduce a Lagrange multiplier $\lambda(\alpha^{n-1}\beta-s^{n-1}\gamma)$ into the mean-field Hamiltonian to include the fluctuations.

According to Eq.~(\ref{eq43}), there is a redundancy for choosing the mean-field parameters.
Furthermore, as mentioned above, $\sigma_{all}^{i+1}\sigma_{aid}^{n-i}$ and $\tau_{all}^{i+1}\tau_{aid}^{n-i}$ have the same commutation relations as the C\&S matrices, and $\sigma_0$ and $\tau_0$ in Eq.~(\ref{eq10}) can be achieved by the projection: $\sigma_0=\sigma_{all}^{i+1}\sigma_{aid}^{n-i}P$ and $\tau_0=\tau_{all}^{i+1}\tau_{aid}^{n-i}P$. Thus, at the mean-field level, we can alternatively represent C\&S matrices by any combinations of $\sigma_{all}^{i+1}\sigma_{aid}^{n-i}$ and $\tau_{all}^{i+1}\tau_{aid}^{n-i}$. Ultimately, the same physical results are obtained by the projection to the physical subspace as well. In the following, we take Baxter's clock model\cite{Baxter1989} and the generalized Kitaev model\cite{Barkeshli2015} as examples to illustrate the applications of the parafermion mean-field theory.

Firstly, we focus on Baxter's clock Hamiltonian\cite{Baxter1989}, because it includes almost all the physics of our interest. It exhibits the similar ordered and disordered phases like the Ising model, and also admits a nonlocal representation wherein the symmetry breaking state is mapped onto a topological phase supporting localized zero modes, i.e., the parafermion zero modes.
The most general nearest-neighbor Hamiltonian for the three-state clock model on an uniform open $L$-site chain is
\begin{equation}
  H_{Z_3}=-J\sum_{a=1}^{L-1}\sigma_a\sigma_{a+1}^\dagger-f\sum_{a=1}^L\tau_a+h.c,
  \label{eq45}
\end{equation}
where $\sigma$ and $\tau$ are the C\&S matrices of order $3$. We represent the spin operators in terms of parafermions:
\begin{equation}
  \sigma_a=\beta_a\gamma_a^{-1},~\sigma_a^\dagger=\alpha_a^{-1}s_a
  ,~\tau_a=\alpha_a^{-1}\beta_a.
  \label{eq46}
\end{equation}
The parafermion operators are ordered along the chain: $\zeta_a\zeta_b=\omega^{{\mathfrak{sgn}}(a-b)}\zeta_b\zeta_a$, and the local order is $s,\alpha,\beta,\gamma$. The terms in the Hamiltonian become
\begin{eqnarray}
  \sigma_a\sigma_{a+1}^\dagger &=& \omega(\gamma_a^{-1}s_{a+1})(\beta_a\alpha_{a+1}^{-1}), \\
  \tau_a &=& \alpha_a^{-1}\beta_a.
\end{eqnarray}
It is easy to verify that $u_{a,a+1}=\gamma_a^{-1}s_{a+1}$ commute with each other and the Hamiltonian: $[u_{a,a+1},u_{b,b+1}]=0$, $[u_{a,a+1}, H_{Z_3}]=0$. In this case, the Hamiltonian is written as
\begin{equation}
    H_{Z_3}=-J\sum_{a=1}^{L-1}\omega u_{a,a+1}\beta_a\alpha_{a+1}^{-1}-f\sum_{a=1}^L\alpha_a^{-1}\beta_a+h.c.,
  \label{eq47}
\end{equation}
where $u_{a,a+1}$ can be considered as a $Z_3$ gauge field, since $u_{a,a+1}^3=1$. Thus, it can be replaced by its eigenvalues. Then, after the projection, the effective Hamiltonian can be considered as a $1$-D parafermion hopping chain. This Hamiltonian is equivalent to that rewritten by the Fradkin-Kadanoff mapping\cite{Fradkin1980}.

In the parafermion representation, the local Hilbert space has been extended from 3-D to 9-D. The redundancy could be projected out by the projector $P$.
If omitting `$h.c.$' terms, the remaining non-Hermitian Hamiltonian can be solved by the notion of free parafermions\cite{Fendley2014}, and there are edge parafermion zero modes in the nontrivial state when $J>f$. The nontrivial state corresponds to the ordered phase of the original spin model, where the ground state has three-fold degeneracy which is topologically protected by the $Z_3$ symmetry. The same results can also be obtained by a mean-field theory. The mean-field Hamiltonian is
\begin{eqnarray}
    H_{\rm mean}=&&-J\sum_{a=1}^{L-1}\omega (u_{a,a+1}\beta_a\alpha_{a+1}^{-1}+u_{a,a+1}^0\gamma_a^{-1}s_{a+1}\nonumber\\
    &&-u_{a,a+1}u_{a,a+1}^0))-f\sum_{a=1}^L\alpha_a^{-1}\beta_a+h.c.,
  \label{eq48}
\end{eqnarray}
where $u_{a,a+1}=\langle\gamma_a^{-1}s_{a+1}\rangle$ and $u_{a,a+1}^0=\langle\beta_a\alpha_{a+1}^{-1}\rangle$. Obviously, it describes the same physics as Eq.~(\ref{eq47}).

\begin{figure}
  \centering
  \includegraphics[width=0.3\textwidth]{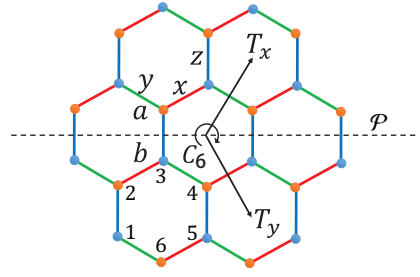}
  \caption{\label{fig}(Color online) Space transformations of the honeycomb lattice. The three bond types are colored differently.}
\end{figure}

Then, coupling the chains together into a honeycomb lattice, we obtain the generalized Kitaev model\cite{Barkeshli2015}:
\begin{equation}
  H_{2D}=-\sum_{\langle ij\rangle}J_{\zeta}(T^\zeta_iT^{\zeta\dagger}_j+h.c.),
  \label{eq49}
\end{equation}
where $\zeta=x,y,z$ depends on the direction of the link $ij$ as shown in Fig.~\ref{fig}, and
\begin{equation}
  T^x=\sigma,~T^y=\tau,~T^z=\tau^{-1}\sigma^{-1},
  \label{eq410}
\end{equation}
which satisfy
\begin{equation}
  T^xT^y=\omega T^yT^x,~T^yT^z=\omega T^zT^y,~T^zT^x=\omega T^xT^z.
  \label{eq411}
\end{equation}
and $T^\zeta_i$ on different sites commute with each other.

Just as before, we write the spin operators in the parafermion representations,
\begin{equation}
  T^x=s^{-1}\alpha,~T^y=s^{-1}\gamma,~T^z=\beta^{-1}s.
\end{equation}
The order of the parafermions is along one chain after another. The local order is different for different sublattices, $s,\alpha,\beta,\gamma$ for $a$ sublattices and $\alpha,\beta,\gamma,s$ for $b$ sublattices. With this representation, the Hamiltonian is written as
\begin{equation}
  H_{2D}=-\sum_{\langle ij\rangle}J_{\zeta}[\omega(\zeta_i\zeta^{-1}_j)(s^{-1}_is_j)+h.c.].
  \label{eq412}
\end{equation}
It is straightforward to verify that $[(\zeta_i\zeta^{-1}_j),(s^{-1}_is_j)]=0$, but they do not commute when the bonds share exactly one site. Thus, there is no exact solution like the Kitaev model with this representation. At the mean-field level, the Hamiltonian is,
\begin{align}
H_{2D}=&-\sum_{\langle ij\rangle}J_{\zeta}[\omega\langle\chi^\zeta_{ij}\rangle s^{-1}_is_j+\omega\langle\chi^0_{ij}\rangle\zeta_i\zeta^{-1}_j \nonumber \\
&-\langle\chi^0_{ij}\rangle\langle\chi^\zeta_{ij}\rangle+h.c.],
\label{eq413}
\end{align}
where the mean-field parameters $\langle\chi^\zeta_{ij}\rangle=\zeta_i\zeta_j^{-1}$ and $\langle\chi^0_{ij}\rangle= s^{-1}_is_j$. It describes 2-D parafermions coupled with a $Z_3$ parafermion gauge field, which has an analogous form to the exact description of the slave genons\cite{Barkeshli2015}.
It is expected that this is a stable mean-field theory which captures the physics of the exact ground state.

The mean-field Hamiltonian suggests that the effective degrees of freedom at low energies are described by parafermions. Thus there might be a low-energy effective $Z_3$ lattice gauge theory\cite{Horn1979,Celik1985} describing the parafermion spin-liquid state as well. To simplify the notations, we set the $Z_3$ vector potential $\sigma_{r,r'}$ and the electric field $\tau_{r,r'}$ only on the nearest-neighbor bonds. The conserved charges are related to the local constraint Eq.~(\ref{eq11})
\begin{equation}
  \alpha^{n-1}\beta\gamma^{n-1}s=\rho_r=\prod_{r'}\tau_{r,r'},
  \label{eq4charge}
\end{equation}
where the product is over the nearest-neighbors $r'$ of the bond $r$. Considering the $Z_3$ gauge field, we have the effective Hamiltonian,
\begin{eqnarray}
  H_{\rm eff}&&=-\sum_{\langle ij\rangle}J_{\zeta}\sigma_{r,r'}[\omega \langle\chi^\zeta_{ij}\rangle s^{-1}_is_j+\omega \langle\chi^0_{ij}\rangle\zeta_i\zeta^{-1}_j+h.c.] \nonumber\\
  &&-\Gamma\sum_{\langle r,r'\rangle}\tau_{r,r'}-\lambda\sum_{p}\prod_{\langle r,r'\rangle\in p}\sigma_{r,r'}+h.c.,
  \label{eq4LGT}
\end{eqnarray}
where $\Gamma,\lambda>0$. The sum is over plaquettes labeled by $p$ in the last term. There will be a variety of phases, here we focus on the deconfined phases in the large $\lambda$ limit or the confined phases in the large $\Gamma$ limit. The physics is simple when $\lambda$ is large enough so that fluctuations of the $Z_3$ flux through the plaquette $\prod_{\langle r,r'\rangle\in p}\sigma_{r,r'}$ are suppressed. The deconfinement is a robust property associated with a gap of the $Z_3$ vortex excitations. In this case, the emergent parafermions carry the $Z_3$ electric charge, which is evident from their minimal coupling to the $Z_3$ gauge field. Like most of states with a deconfined $Z_3$ lattice gauge field, this state can also be characterized by a $Z_3$ topological order. In the other limit when $\Gamma$ dominates, the local constraint is restored and the Hilbert space is projected back to that of the spin model. Consequently, the fractionalized parafermions are confined. To the second-order perturbation, the effective Hamiltonian is given by
\begin{equation}
  H_{\rm eff}^{(2)}=\sum_{\langle ij\rangle}(J_{\zeta,\langle r,r'\rangle}T^\zeta_iT^{\zeta\dagger}_j+h.c.),
  \label{eq4eff}
\end{equation}
where $J_{\zeta,\langle r,r'\rangle}=2( \omega J_\zeta\langle\chi^\zeta_{ij}\rangle\langle\chi^0_{ij}\rangle+h.c.)/(3\Gamma)$. When $J_{\zeta,\langle r,r'\rangle}<0$, it looks like the original spin model in Eq.~(\ref{eq49}). Note that the model and its results can be generalized to any $n$-order C\&S matrices and parafermions.

To solve the parafermion Hamiltonian, we have to seek some unique methods, or require a Fock space\cite{Cobanera2014} of the parefermions to construct the occupation number representation, or adopt the eigenvectors\cite{Wei2015} of the parafermion Hamiltonian. In any case, it is not easy to solve the mean-field parafermion Hamiltonian. Rather, the PSG classification of possible parafermion spin-liquid states is much easier, and it can provide some crucial information to simplify the solution of the mean-field parafermion Hamiltonian. Therefore, in the following, we will focus on the PSG classification of parafermion spin liquids.

\section{Projective implementation of symmetries}\label{PSG}

The PSG classification is introduced by Wen\cite{Wen2002} on the square lattice for the so-called symmetrical liquids, i.e., the spin-liquid states that preserve all of the lattice, spin rotation, parity, and time reversal symmetries (or a subgroup of these symmetries, if we use a looser definition of spin liquids).
The PSG approach has been a successful method for the construction and classification of the quadratic mean-field Hamiltonians of a given lattice spin model. In fact, even beyond the quadratic mean-field Hamiltonians, it also in principle allows to distinguish different phases as long as the symmetry can be projectively realized. Moreover, based on the Majorana-fermion representation, the simplest parafermion representation, the Kitaev model has been shown to have the same PSG classifications as those of the Schwinger-fermion approach for the SU($2$) spin liquids\cite{You2012,Chen2012}. In this section, we will perform a projective realization of the symmetry group (SG) of an SU($n$) spin model in the parafermion mean-field theory.

\subsection{Emergent gauge symmetry}
As mentioned in the sections \ref{parton} and \ref{liquid}, the enlarged parafermion Hilbert space leads to additional local gauge symmetries ${\rm G}=\{\bigotimes_{k\neq0} U_k|U_k\in{\rm SU}(n)\}$, and the different mean-field states related by the gauge transformations give rise to the same physical stats after projection. 
However, in the mean-field Hamiltonian $H_0$, the gauge symmetries are generally not preserved, neither are the physical symmetries of the spin Hamiltonian $H$. This implies an interpretation about the symmetries of the mean-field Hamiltonian.

The additional gauge redundancy in the parafermion space means that there is some freedom how physical symmetries act in the Hilbert space of parafermions. A symmetry transformation $w\in{\rm SG}$ can be restored with an gauge transformation $g\in{\rm G}$:
\begin{eqnarray}
 w(H) &=& H, \nonumber\\
 gw(H_0) &=& H_0.
 \label{eq51}
\end{eqnarray}
Therefore, we can define the symmetry group of $H_0$ as ${\rm PSG}=\{gw|gw(H_0)=H_0,w\in{\rm SG},g\in{\rm G}\}$. However, this choice is not arbitrary, since the gauge transformations must respect the algebraic relations among the symmetrical transformations. Moreover, PSG contains a special subgroup called invariant gauge group (IGG) in which the group elements are the gauge transformations leaving $H_0$ unchanged: ${\rm IGG}=\{g|g(H_0)=H_0,g\in{\rm G}\}$. If we define a linear map from PSG to SG, SG is the image and the IGG is the kernel manifestively. Thus, in the mathematical relationship,
\begin{equation}
  {\rm SG=PSG/IGG},
  \label{eq52}
\end{equation}
we say that the SG is represented (projectively) in the parafermion Hilbert space as PSG. This is the core of the PSG classification and in the following we will introduce the representations of symmetries that act on a single site.

\subsection{Projective realization of the symmetries}

Considering a space transformation $\mathcal{S}:r\rightarrow \mathcal{S}(r)$, to make $H_0$ invariant, the accompanying gauge transformations are required. Then we have the projective transformation on the parafermion matrices written as
\begin{equation}
  \mathcal{S}:P_{kr}^{[i]}\rightarrow U_k^\dagger P_{k\mathcal{S}(r)}^{[i]}U_{k+i},
  \label{eq53}
\end{equation}
where $U_0=1$ and $U_k\in {\rm SU}(n)$ ($k\neq0$), which do not affect the transformations of the gauge-invariant spin operators. Similarly, time reversal transformation is implemented with an antiunitary operation,
\begin{equation}
  \mathcal{T}: P_{kr}^{[i]}\rightarrow U_k^\dagger \mathcal{K} P_{kr}^{[i]}\mathcal{K} U_{k+i},
  \label{eq54}
\end{equation}
where $\mathcal{K}$ is the complex conjugate operator, $U_k\in {\rm SU}(n)$ and $U_0$ is the spin rotation operator. Specially, $U_0$ is the identity operator for the spinless particles and $i\sigma^y$ for the $s=1/2$ SU($2$) spin particles.

Spin rotation transformations $R$ are a bit of difference. If we follow the assumptions proposed by Chen et.al.\cite{Chen2012}, the projective spin rotation operators are
\begin{equation}
  T^{ab}=p_0^{[0]ab}+\sum_{k=1}^{n-1}\epsilon_k p_k^{[0]ab},
  \label{eq55}
\end{equation}
where $\epsilon_k=0,1$. There is no doubt that they have the same commutation relations as $p_0^{[0]ab}$, and $PT^{ab}P=p_0^{[0]ab}$. For the parafermion matrices, they are written as
\begin{equation}
  R:P_{kr}^{[i]}\rightarrow U_k^\dagger P_{kr}^{[i]}U_{k+i},
  \label{eq56}
\end{equation}
where $U_k=I$ if $\epsilon_k=0$, otherwise $U_k=U_0$, which are the physical spin rotation transformations. There is also a special conserved operator associated with each plaquette, for example that defined on the Kitaev model\cite{Kitaev2006,Kells2008,Barkeshli2015}, which can also be regard as a product of a loop of spin rotation operators around plaquettes. Thus, it should be projectively realized like the spin rotation transformations.

The actions of the transformations of IGG on the parafermion matrices are given by
\begin{equation}
  g:P_{kr}^{[i]}\rightarrow U_k^\dagger P_{kr}^{[i]}U_{k+i},
  \label{eq57}
\end{equation}
where again $U_k(k\neq0)$ must be chosen to leave $H_0$ invariant. If relating IGG to a symmetry transformation, the associated transformation is the identity transformation.

PSGs are often classified by the type of IGG. For example, if ${\rm IGG}=Z_2$, we say that the PSG is a $Z_2$ PSG. It should be remarked that the PSG classification in the parafermion parton approach is very similar to that in the SU($2$) case, and the main difference is that the gauge structure for parafermion partons is $\{\bigotimes{\rm SU}(n)\}^{n-1}$ rather than SU($2$). Thus, the IGG is the subgroup of $\{\bigotimes{\rm SU}(n)\}^{n-1}$, and the spin liquids are named by the IGG. However, it is important to keep in mind that the classification of PSGs is not the same as a classification of spin-liquid states. There can be distinct spin liquids with the same PSG, so our aim is to determine the number of PSG classes and how many different choices in each PSG class. 
To recognize the PSG of the generalized Kitaev model of order $3$, in section \ref{Z3}, we will give the classification of the $Z_3$ PSGs on the honeycomb lattice as an example.

\subsection{Symmetry of the generalized Kitaev model}

Now we will consider the symmetry group of the generalized Kitaev model. Manifestly, the space group is generated by two translations $T_x$ and $T_y$, and a $6$-fold rotation $C_6$ accompanied by a spin rotation $\bm{S}_{C_6}$. Unlike the Kitaev model, the parity $\mathcal{P}$ (which is chosen to be the reflection along the $z$-bond in this paper) and time reversal $\mathcal{T}$ symmetries are absent for the generalized Kitaev model. However, their combination transformation $\mathcal{P}\mathcal{T}$ is invariant with a spin rotation $\bm{S}_{\mathcal{P}\mathcal{T}}$. $\bm{S}_{C_6}$ and $\bm{S}_{\mathcal{P}\mathcal{T}}$ read
\begin{eqnarray}
  \bm{S}_{C_6}&=&{e^{i\pi /18}\over \sqrt{3}}(T^x+T^y+T^z),\nonumber\\
  \bm{S}_{\mathcal{P}\mathcal{T}}&=&{e^{-i\pi /6}\over 3i}\sum_{a,b}\omega^{a\neq b}\sigma^a\tau^b.
  \label{eq414}
\end{eqnarray}
Specially, following the Kitaev model, a similar loop conserved operator on each hexagonal plaquette is defined as $W_p=T_1^xT_2^{y\dagger}T_3^zT_4^{x\dagger}T_5^yT_6^{z\dagger}$, where the site labels are shown in Fig.~\ref{fig}. We can easily verify that $W_p^3=1$, and $[W_p,H_{2D}]=0$, so the Hilbert space can be separated according to the eigenstates of $W_p$. When mapping to the parafermion Hamiltonian, this loop operator can be defined as a Wilson loop $W_p=\chi_{12}^0\chi_{12}^z\chi_{23}^0\chi_{23}^x\chi_{34}^{0\dagger}\chi_{34}^{y\dagger}\chi_{45}^{0\dagger}\chi_{45}^{z\dagger}\chi_{56}^{0\dagger}\chi_{56}^{x\dagger}\chi_{61}^0\chi_{61}^y$, whose phase can be regard as a flux through the plaquette.

Written in the parafermion representation, the C\&S matrices are
\begin{eqnarray}
  \sigma&=&\beta\gamma^{-1}+s^{-1}\alpha+s\alpha^{-1}\beta^{-1}\gamma,\nonumber\\
  \tau&=&\alpha^{-1}\beta+s^{-1}\gamma+s\alpha\beta^{-1}\gamma^{-1}.
  \label{eq415}
\end{eqnarray}
The spin operators are decomposed into the three parafermion matrices $P_0^{[1]}$, $P_1^{[1]}$ and $P_2^{[1]}$ according to Eq.~(\ref{eq17}). Then there are three SU($3$) transformations $D$, $Z$ and $Y$, under which $P_0^{[1]}$, $P_1^{[1]}$ and $P_2^{[1]}$ are transformed as $D^\dagger P_0^{[1]}Z$, $Z^\dagger P_1^{[1]}Y$, and $Y^\dagger P_2^{[1]}D$. The transformations $Z$ and $Y$ are the two gauge transformations that leave the spin operators invariant, and $D=S_{\mathcal{P}\mathcal{T}}S^*S_{\mathcal{P}\mathcal{T}}^\dagger$ is related to a spin rotation. Here, $S$ is the matrix representation of a spin rotation, and $S_{\mathcal{P}\mathcal{T}}$ is the matrix representation of the accompanying spin rotation of $\mathcal{P}\mathcal{T}$ transformation.

Any operator acting on a spin, such as a symmetric transformation, also acts on the gauge space. Thus, when we fractionalize spins in the parafermion representation, we must also specify how the symmetrical operations of the model act on the gauge degree of freedom. This extra information, known as the projective symmetry group\cite{Wen2002}, characterizes the fractionalized quantum state. If the ground state is characterized by Eq.~(\ref{eq413}), where the parafermion $s$ is connected while the others are isolated in different bonds just like that in the SU($2$) case\cite{You2012}, the PSG of parafermion spin liquid can be determined starting from the fact that the parafermion $s$ is a special flavor which should not be mixed with others. Thus, any symmetrical transformation must preserve the flavor $s$ with an accompanying gauge transformation, otherwise the symmetry will break in the ground state. Under the transformation $D=Z=Y$, $s$ is invariant, while $\alpha$ ($\gamma$) varies like $\sigma$ ($\tau$) under a SU($3$) transformation. Therefore, the accompanying gauge transformation for one of the symmetrical transformation $D$ is required to be $Z=\omega^mD$ and $Y=\omega^nD$, where $m$ and $n$ are determined by keeping $\langle\chi_{ij}^\zeta\rangle$ and $\langle\chi_{ij}^0\rangle$ unchanged.

The local gauge symmetry of the generalized Kitaev model is ${\rm SU(3)}\otimes{\rm SU(3)}$, which will degenerate into an IGG in the mean-field Hamiltonian. Since the mean-field Hamiltonian Eq.~(\ref{eq413}) is definitely invariant under the transformations $\zeta\rightarrow\omega^n\zeta$ and $s\rightarrow\omega^ns$, IGG is at least a $Z_3$ group $\{1, e^{i2\pi/3}\otimes e^{-i2\pi/3},e^{-i2\pi/3}\otimes e^{i2\pi/3}\}$. For the mean-field Hamiltonian, we suppose that the IGG is $Z_3$ and the flux through each hexagon is $\Phi_p$. Because $S_{C_6}^6=1$ and $S_{C_6}S_{\mathcal{P}\mathcal{T}}S_{C_6}^*S_{\mathcal{P}\mathcal{T}}^*=1$, according to the results in the following section, a preliminary analysis suggests that the PSG belongs to the class $\Phi_p$(I), where $\Phi_p$ denotes the flux through each hexagon in the mean-field Hamiltonian, and (I) denotes the constraint $\eta_{C_6}\eta_{xy}^2=1$.

\section{Classification of $Z_3$ PROJECTIVE SYMMETRY GROUPS ON A HONEYCOMB LATTICE}\label{Z3}

Here, we present the classification of $Z_3$ PSG of a SU(3) spin model on a Honeycomb lattice with SG generated by $\{T_x,T_y,C_6,\mathcal{P}\mathcal{T}(\mathcal{P},\mathcal{T})\}$. $T_x$ and $T_y$ are two translations in the two dimensions, $C_6$ is a $6$-fold rotation, and $\mathcal{P}$ is a reflection, as illustrated in Fig.~\ref{fig}. $\mathcal{T}$ is the time-reversal transformation. Considering that the SU($3$) parafermion representation has two SU($3$) gauge transformations, we suppose the $Z_3$ IGG is $\{1, e^{i2\pi/3}\otimes e^{-i2\pi/3},e^{-i2\pi/3}\otimes e^{i2\pi/3}\}$.

The symmetry group of a general spin model on the Honeycomb lattice generated by the three generators $T_x,T_y,C_6$ has the following four relations,
\begin{align}
  C_6^6&=1, \nonumber \\
  T_xT_yT_x^{-1}T_y^{-1}&=1, \nonumber\\
  C_6T_xC_6^{-1}T_x^{-1}T_y^{-1}&=C_6T_yC_6^{-1}T_x=1.
  \label{eq61}
\end{align}
When $\mathcal{P}\mathcal{T}$ presents, there are extra four relations,
\begin{eqnarray}
  \mathcal{P}\mathcal{T}\mathcal{P}\mathcal{T} &=& 1, \nonumber\\
  \mathcal{P} \mathcal{T}T_x\mathcal{P}\mathcal{T}T_y^{-1} &=& \mathcal{P}\mathcal{T}T_y\mathcal{P}\mathcal{T}T_x^{-1}=1, \nonumber\\
  C_6\mathcal{P}\mathcal{T} C_6 \mathcal{P}\mathcal{T} &=& 1,
  \label{eq61x}
\end{eqnarray}
Furthermore, if both $\mathcal{P}$ and $\mathcal{T}$ present, nine extra relations will arise instead,
\begin{eqnarray}
  \mathcal{T}^2&=&\mathcal{P}^2= 1, \nonumber\\
  \mathcal{P}\mathcal{T}\mathcal{P}\mathcal{T} &=& 1, \nonumber\\
  \mathcal{T}T_x\mathcal{T}T_x^{-1} &=&\mathcal{T}T_y\mathcal{T}T_y^{-1}= 1, \nonumber\\
  \mathcal{P} T_x\mathcal{P}^{-1}T_y^{-1} &=& \mathcal{P} T_y\mathcal{P}^{-1}T_x^{-1}=1, \nonumber\\
  C_6\mathcal{P} C_6 \mathcal{P} &=& 1, \nonumber\\
  \mathcal{T}C_6\mathcal{T}C_6^{-1} &=& 1.
  \label{eq61y}
\end{eqnarray}

The identity transformations are defined from these relations, and their accompanying gauge transformations are given by $\{\mathrm{IGG}|\eta_{ab}\in \mathrm{IGG}\}$, where $a$ and $b$ indicate the corresponding symmetries (see Appendix A for details). For each relation, there are three different classes of solutions. Similar but more complicated results are obtained for all relations between the symmetries. Correspondingly, different combinations of $\eta_{ab}$ finally define distinct characteristics of spin-liquid states, yielding a classification scheme on a general ground. The detailed derivation is given in Appendix A, and the corresponding results are presented in the following.

First, if only two translations are involved, fixing the relative gauge between unit cells: $g_{T_y}(x,y)=g_{T_x}(x,0)=1$, the solution for the two translations are
\begin{equation}
  g_{T_x}(x,y)=\eta_{xy}^y,~g_{T_y}(x,y)=1.
  \label{eq62}
\end{equation}
Since $\eta_{xy}\in{\rm IGG}$, under PSG classification, there are only three types of $Z_3$ spin liquids which can be determined by the flux $\Phi_p=2k\pi/3$ through each hexagon in the mean-field Hamiltonian without other symmetries.

Then, including the $C_6$ rotation, the accompanying gauge transformation of $C_6$ satisfies
\begin{align}
g_{C_6}(x,y)=\eta_{xy}^{x(x-1)/2-xy}g_{C_6}(0,0),
\end{align}
where $(0,0)$ will be omitted in the following. There are further three new types of classifications for each translation symmetry, depending on the value of $\eta_{C_6}\eta_{xy}^2$. Thus, there are $3\times3=9$ types of the classifications in total. In the following, we discuss the solutions of $g_{C_6}(a)$ and $g_{C_6}(b)$ ($a$ and $b$ refer to different sublattices) in three different situations according to the values of  $\eta_{C_6}\eta_{xy}^2$.

(I): $\eta_{C_6}\eta_{xy}^2=1$. $g_{C_6}(a)$ and $g_{C_6}(b)$ have $4\times4=16$ different choices (The accompanying gauge transformations $g_\alpha$ are elements of SU($3$)$\otimes$SU($3$) and $1$ is short for $1\otimes1$, unless otherwise specified). Given a fixed global gauge, we have
\begin{equation}
  g_{C_6}(a)=1,~g_{C_6}(b)=m_+\otimes m_-,
  \label{eq63}
\end{equation}
where $m_+$ and $m_-$ are elements of the group $\{1,{\rm exp}(i2\pi/3),{\rm exp}(-i2\pi/3),{\rm exp}(i2\pi\lambda_3/3)\}$. Here, $\lambda_3$ is the third Gell-Mann matrix
\begin{equation}
  \lambda_3=\left(
  \begin{array}{ccc}
    1 & 0 & 0 \\
    0 & -1 & 0 \\
    0 & 0 & 0
  \end{array}
  \right).
  \nonumber
\end{equation}

(II) and (III): $\eta_{C_6}\eta_{xy}^2={\rm exp}(\xi i2\pi/3)\otimes{\rm exp}(-\xi i2\pi/3)$, where $\xi$ takes $+$ and $-$ for (II) and (III), respectively. Each type has $3\times3=9$ different choices. $g_{C_6}(a)$ and $g_{C_6}(b)$ have the same form as that in Eq.~(\ref{eq63}), but
\begin{equation}
  m_\pm\in\{{\rm exp}\left[\xi i2\sqrt{3}\pi \lambda_8({1\over 3}k\pm{1\over 9}) \right]|k=0,1,2\},
  \nonumber
\end{equation}
where $\lambda_8$ is the eighth Gell-Mann matrix
\begin{equation}
 \lambda_8={1\over \sqrt{3}}\left(
  \begin{array}{ccc}
    1 & 0 & 0 \\
    0 & 1 & 0 \\
    0 & 0 & -2
  \end{array}
  \right).
  \nonumber
\end{equation}

Therefore, in all, there are $9$ types and $3\times(16+9+9)=102$ different choices.

The relations in Eq.~(\ref{eq61x}) involve $\mathcal{P}\mathcal{T}$ result $\eta_{\mathcal{P}\mathcal{T}}=\eta_{x\mathcal{P}\mathcal{T}}=\eta_{y\mathcal{P}\mathcal{T}}=\eta_{C_6\mathcal{P}\mathcal{T}}=1$, which means there is no extra gauge freedom for $\mathcal{P}\mathcal{T}$. Therefore, with a fixed global gauge, we have
\begin{equation}
  g_{\mathcal{P}\mathcal{T}}(x,y,a)=g_{\mathcal{P}\mathcal{T}}(x,y,b)=\eta_{xy}^{xy},
  \label{eq6PT}
\end{equation}
and the classification is the same as that with the symmetries $T_x$, $T_y$, and $C_6$.

Finally, taking both $\mathcal{P}$ and $\mathcal{T}$ symmetries into account instead of $\mathcal{P}\mathcal{T}$ symmetry, there are $3\times3=9$ types of classifications based on the values of $\eta_\mathcal{P}$ and $\eta_{C_6\mathcal{P}}$. When $\mathcal{P}$ or $\mathcal{T}$ presents, $\eta_{xy}=1$. If $\eta_{xy}\neq1$, there is $2\pi/3$ ($4\pi/3$) flux through each hexagonal plaquette in the mean-field Hamiltonian. Since the parity (which is also the reflection here) and time-reversal transformations would reverse the flux, the Hamiltonian is inevitable to vary according to these transformations. Thus, it is impossible to realize $\mathcal{P}$ and $\mathcal{T}$ projectively when $\eta_{xy}\neq1$. After a careful scrutiny, we can get $\eta_{C_6}=\eta_\mathcal{T}=1$ and $\eta_{\mathcal{T}C_6}\eta_\mathcal{P}=\eta_{\mathcal{T}\mathcal{P}}\eta_\mathcal{P}^2\eta_{C_6\mathcal{P}}^2=1$. Consequently, the accompanying gauge transformations of $\mathcal{P}$ and $\mathcal{T}$ are translational invariant, i.e., $g_\mathcal{P}(x,y)=g_\mathcal{P}$ and $g_\mathcal{T}(x,y)=g_\mathcal{T}$. Therefore, with a fixed global gauge, the accompanying gauge transformations can be chosen as
\begin{eqnarray}
  g_{T_x}=&g_{T_y}&=1, \nonumber \\
  g_{C_6}(a)&=&1, \nonumber \\
  g_{C_6}(b)&=&\eta_\mathcal{P}^2\eta_{C_6\mathcal{P}}, \nonumber \\
  g_\mathcal{P}(b) &=&m_+\otimes m_-  \nonumber\\
  g_\mathcal{P}(a) &=& \eta_\mathcal{P}\eta_{C_6\mathcal{P}}^2g_\mathcal{P}(b) \nonumber\\
  g_\mathcal{T}(a) &=&1, \nonumber \\
  g_\mathcal{T}(b)&=& \eta_{C_6\mathcal{P}}^2\eta_\mathcal{P},
  \label{eq64}
\end{eqnarray}
where $m_\pm\in\{{\rm exp}\left[i2\sqrt{3}\pi \lambda_8(k/2\pm l/6)\right]|k=0,1\}$
and $l=0,1,2$ for three different choices of $\eta_{C_6\mathcal{P}}$. With the values of $k$, there are $2\times2=4$ different choices for each type. Thus, there are $9\times4=36$ different choices in all. In appendix, we give an extra discussion on the time-reversal and reflection symmetry. If the time-reversal symmetry is absent, the classification is generally the same, but, if the reflection symmetry is absent, there will be three PSGs related to three different values of $\eta_{\mathcal{T}C_6}$, and each type has only one solution.

\section{DISCUSESion and Conclusion}

In this paper, we develop a theoretical framework to describe spin-liquid states based on the parafermion parton approach and effective gauge theory.
We show that SU($n$) spins can be fractionalized appropriately by the parafermion operators with $n-1$ SU($n$) gauge symmetries, which are distinct from the fermions and bosons in the conventional condensed matters. By this parton approach, the symmetries of an original SU($n$) spin model can be projectively realized in the effective mean-field Hamiltonian.

As a concrete example, we have discussed the PSG classifications of the parafermion spin liquids of the generalized Kitaev model. The projective realization of this model leads to spin liquids with parafermion excitations coupled to a deconfined $Z_3$ gauge field. Although the local operators of this example are $Z_3$ rotors rather than real SU($3$) spins, the parafermion parton approach and PSG classification work also for SU($n$) spin models.
The local gauge symmetry of the $Z_3$ PSGs discussed here is SU($3$)$\otimes$SU($3$), which is much more complicated than that in the SU($2$) models with spin $s=1/2$, whose IGGs are always $Z_2$, U($1$) and SU($2$). Other IGGs may lead to more kinds of the parafermion spin liquids, which is an interesting open question in the future.
Beyond the classification, it is difficult to diagonalize the parafermion Hamiltonians actually, since there is no general method as what we have for the quadratic fermion Hamiltonians.
At the present stage, we can of course technically use the numerical method based on the standard fermion parton to study SU($N$) spin models\cite{PhysRevB.86.224409,PhysRevB.86.224409,PhysRevLett.117.167202}, especially when the ground states are related to the fermionic fractionization. However, the fermion representation is limited for parafermion spin liquids. When $N=2$, the fermion representation is equivalent to the Majorana fermions representation, and when $N=3$, it may be feasible since the projection may lead the fermion mean-field wave function to the true ground state. But when $N$ increases to large values, fermions differ significantly from parafermions, it is hard to expect the fermion mean-field wave function could lead to a parafermion spin liquid state. Now the numerical methods about parafermion states is developing, such as the matrix product states for parafermions introduced recently\cite{PhysRevB.95.195122}.
Thus, in order to understand the parafermion spin liquids much clearer, not only deeper analysis but also more numerical calculation methods are needed, and our parton approach deserve further attention.

Apart from the pure theoretical significance that it provides a new method to study quantum spin liquids, it is also conducive to the experiments on the SU($N$) magnets \cite{Gorshkov2010,Pagano2014,Scazza2014,XZhang2014,Cazalilla2014,Onufriev1999,KarloPenc2003} to explore the exotic parafermion excitations. These experimental platforms will give rise to the topological qubits, which are better protected against environmental noise and allow for richer fault-tolerant qubit rotations in contrast to the Majorana-based architectures.

\begin{acknowledgments}
This work was supported by the National Natural Science Foundation of China (11374138, 11674158 and 11774152)
and National Key Projects for Research and Development of China (Grant No. 2016YFA0300401).
\end{acknowledgments}

\appendix
\section{DERIVATION OF PSG ON A HONEYCOMB LATTICE}
On the Honeycomb lattice, each unit cell is labeled by its integer coordinates $x$ and $y$ along the translation axes of $T_x$ and $T_y$. A spin site is further specified by its sublattice label $a$ or $b$ within the unit cell as shown in Fig.~\ref{fig}. The symmetry group operators act on the lattice by
\begin{eqnarray*}
  T_x(x,y) &=& (x+1,y), \\
  T_y(x,y) &=& (x,y+1), \\
  C_6(x,y,a) &=& (x-y,x,b), \\
  C_6(x,y,b) &=& (x-y-1,x,a), \\
  \mathcal{P}(x,y,\zeta) &=&(y,x,\overline{\zeta}).
\end{eqnarray*}
The sublattice label is omitted if a formula applies to both sublattices. The reflection transformation does not only exchange the $x$ and $y$ but also the sublattice labels $a$ and $b$.

According to the relations of symmetrical transformations in Eq.~(\ref{eq61}), (\ref{eq61x}) and (\ref{eq61y}), the constraints of the accompanying gauge transformations of $T_x$, $T_y$ and $C_6$ are
\begin{eqnarray}
  g_{C_6}(C_6^3(r))g_{C_6}(C_6^2(r))g_{C_6}(C_6(r))&&\nonumber\\
  g_{C_6}(r)g_{C_6}(C_6^{-1}(r))g_{C_6}(C_6^{-2}(r))&=&\eta_{C_6},\label{eqA9}\\
  g_{T_x}(T_y\circ T_x(r))g_{T_y}(T_x(r))&& \nonumber\\
  g_{T_x}^{-1}(r)g_{T_y}^{-1}(T_y(r)) &=& \eta_{xy}, \label{eqA1}\\
  g_{C_6}(T_x\circ C_6^{-1}(r))g_{T_x}(C_6^{-1}(r))&& \nonumber \\
  g_{C_6}^{-1}(r)g_{T_x}^{-1}(T_x(r))g_{T_y}^{-1}(T_y\circ T_x(r))  &=&\eta_{C_6x}, \label{eqA4}\\
  g_{C_6}(T_y (r))g_{T_y}(r)&& \nonumber \\
  g_{C_6}^{-1}(C_6(r))g_{T_x}(T_x^{-1}\circ C_6(r))&=&\eta_{C_6y}.\label{eqA5}
\end{eqnarray}
 When involving $\mathcal{P}\mathcal{T}$, the additional constraints of the accompanying gauge transformations are
\begin{eqnarray}
  g_\mathcal{\mathcal{P} T}(\mathcal{P}(r))\mathcal{K}g_{\mathcal{P}\mathcal{T}}(r)\mathcal{K} &=&\eta_{\mathcal{\mathcal{P} T}}, \label{eqAx1}\\
  g_{\mathcal{P}\mathcal{T}}(T_x(r))\mathcal{K} g_{T_x}(r)g_{\mathcal{P}\mathcal{T}}(\mathcal{P}(r))\mathcal{K}&&\nonumber\\
  g_{T_y^{-1}}(T_y\circ\mathcal{P}(r)) &= &\eta_{x\mathcal{P}\mathcal{T}}, \label{eqAx2}\\
  g_\mathcal{P}\mathcal{T}(T_y(r))\mathcal{K} g_{T_y}(r)g_{\mathcal{P}\mathcal{T}}(\mathcal{P}(r))\mathcal{K}&&\nonumber\\
  g_{T_x^{-1}}(T_x\circ\mathcal{P}(r)) &= &\eta_{y\mathcal{P}\mathcal{T} }, \label{eqAx3}\\
  g_{C_6}(\mathcal{P}\circ C_6(r))g_{\mathcal{P}\mathcal{T}}(C_6(r))&&\nonumber\\
  g_{C_6}(r) g_{\mathcal{P}\mathcal{T}}(\mathcal{P}(r)) &=& \eta_{C_6\mathcal{P}\mathcal{T}}. \label{eqAx4}
\end{eqnarray}
If involving both $\mathcal{P}$ and $\mathcal{T}$ instead of $\mathcal{P}\mathcal{T}$, they are
\begin{eqnarray}
  g_\mathcal{T}(T_x(r))g_{T_x}(r)&&\nonumber\\
  g_\mathcal{T}(r)g_{T_x}^{-1}(T_x(r)) &=& \eta_{x\mathcal{T}},\label{eqA2}\\
  g_\mathcal{T}(T_y(r))g_{T_y}(r)&&\nonumber\\
  g_\mathcal{T}(r)g_{T_y}^{-1}(T_y(r)) &=& \eta_{y\mathcal{T}},\label{eqA3}\\
  g_\mathcal{P}(T_x(r)) g_{T_x}(r)g_\mathcal{P}^{-1}(\mathcal{P}(r))&&\nonumber\\
  g_{T_y^{-1}}(T_y\circ\mathcal{P}(r)) &= &\eta_{\mathcal{P} x}, \label{eqA6}\\
  g_\mathcal{P}(T_y(r)) g_{T_y}(r)g_\mathcal{P}^{-1}(\mathcal{P}(r))&&\nonumber\\
  g_{T_x^{-1}}(T_x\circ\mathcal{P}(r)) &= &\eta_{\mathcal{P} y}, \label{eqA7}\\
  g_\mathcal{T}(r)\mathcal{K}g_\mathcal{T}(r)\mathcal{K} &=&\eta_\mathcal{T},\label{eqA8}\\
  g_\mathcal{P}(\mathcal{P}(r))g_\mathcal{P}(r)&=& \eta_{\mathcal{P}}, \label{eqA10}\\
  g_\mathcal{T}(C_6(r))\mathcal{K}g_{C_6}(r)g_\mathcal{T}(r)\mathcal{K}g_{C_6}^{-1}(C_6(r)) &=&\eta_{\mathcal{T}C_6}, \label{eqA11}\\
  g_\mathcal{T}(\mathcal{P}(r))\mathcal{K}g_\mathcal{P}(r)g_\mathcal{T}(r)\mathcal{K}g_\mathcal{P}^{-1}(\mathcal{P}(r)) &=&\eta_{\mathcal{T}\mathcal{P}}, \label{eqA12}\\
  g_{C_6}(\mathcal{P}\circ C_6(r))g_\mathcal{P}(C_6(r))&&\nonumber\\
  g_{C_6}(r) g_\mathcal{P}(\mathcal{P}(r)) &=& \eta_{C_6\mathcal{P}}. \label{eqA13}
\end{eqnarray}
Where $\eta_{ab}$ are the group elements of the IGG. It is noticed that $1$ and $e^{\pm i2\pi/3}$ in the elements of IGG are the SU($3$) matrices in the form of ${\rm exp}(i2\sqrt{3}\lambda_8k/3)$, rather than a scalar. Thus, the calculations involving them must obey the SU($3$) algebra. Although there are so many conditions, not all of them are independent. Therefore, it is not a simple classification including $3^{8}$ types of PSG in the case that there is only $\mathcal{P}\mathcal{T}$ symmetry or $3^{13}$ types in the case that there are both $\mathcal{P}$ and $\mathcal{T}$ symmetries.

It is known that the wave functions are gauge equivalence if they can be related by the local gauge transformations. Making use of such degrees of freedom, we can always choose relative gauge between the unit cells so that
\begin{equation}
  g_{T_x}(x,0)=g_{T_y}(x,y)=1.
  \label{eqA14}
\end{equation}
Now, according to Eq.~(\ref{eqA1}), we have $g_{T_x}(x,y+1)=\eta_{xy}g_{T_x}(x,y)$. The solutions are given by
\begin{equation}
  g_{T_x}(x,y)=\eta_{xy}^y,~g_{T_y}(x,y)=1.
  \label{eqA15}
\end{equation}

Substituting Eq.~(\ref{eqA15}) into Eq.~(\ref{eqA4}) and (\ref{eqA5}), we have
\begin{eqnarray}
  g_{C_6}(x+1,y) &=& \eta_{xy}^{x-y}\eta_{C_6x}g_{C_6}(x,y), \nonumber\\
  g_{C_6}(x,y+1) &=& \eta_{xy}^{-y}\eta_{C_6y}g_{C_6}(x,y),
  \label{eqA16}
\end{eqnarray}
whose solution is
\begin{equation}
  g_{C_6}(x,y)=\eta_{xy}^{x(x-1)/2-xy}\eta_{C_6x}^x\eta_{C_6y}^yg_{C_6}.
  \label{eqA17}
\end{equation}
Because $g_{T_x}$ or $g_{T_y}$ appears only once in Eqs.~(\ref{eqA4}) and (\ref{eqA5}), if we choose a gauge transformation of $g_{T_x}$ or $g_{T_y}$ by multiplying it by an element of IGG, the solution of $\eta_{C_6x}$ or $\eta_{C_6y}$ will be changed. However the mean field Hamiltonian is invariant, since each term of it is degree $[0]$. Therefore, $g_{T_x}$ and $g_{T_y}$ (also $\eta_{C_6x}$ and $\eta_{C_6y}$) are not independent. Making use of this gauge freedom, we can set $\eta_{C_6x}=\eta_{C_6y}=1$. Then, inserting Eq.~(\ref{eqA17}) into Eq.~(\ref{eqA9}), we have
\begin{equation}
  \left[g_{C_6}(a)g_{C_6}(b)\right]^3=\left[g_{C_6}(b)g_{C_6}(a)\right]^3=\eta_{C_6}\eta_{xy}^2.
  \label{eqA18}
\end{equation}
It is clear that the solutions are based on the value of $\eta_{C_6}\eta_{xy}^2$, which is surely one element in IGG. Fixing the relative gauge between sublattices $a$ and $b$ by supposing $g_{C_6}(a)=1$, we have $g_{C_6}(b)^3=\eta_{C_6}\eta_{xy}^2$. According to its solutions, there are three types of PSG as mentioned in the section \ref{Z3}.

Next, we are going to calculate PSGs, in the cases that there is $\mathcal{P}\mathcal{T}$ and there are both $\mathcal{P}$ and $\mathcal{T}$.

\subsection{$\mathcal{P}\mathcal{T}$}
When $\mathcal{P}\mathcal{T}$ presents, there are additional constraints Eq.~(\ref{eqAx1}), (\ref{eqAx2}), (\ref{eqAx3}) and (\ref{eqAx4}). Substituting Eq.~(\ref{eqA14}) into Eq.~(\ref{eqAx2}) and (\ref{eqAx3}), we obtian
\begin{eqnarray}
 g_{\mathcal{P}\mathcal{T}}(y+1,x)\mathcal{K}g_{\mathcal{P}\mathcal{T}}(x,y)\mathcal{K}&=&\eta_{xy}^y\eta_{x\mathcal{P}\mathcal{T}},\nonumber\\
 g_{\mathcal{P}\mathcal{T}}(y,x+1)\mathcal{K}g_{\mathcal{P}\mathcal{T}}(x,y)\mathcal{K}&=&\eta_{xy}^x\eta_{y\mathcal{P}\mathcal{T}},
  \label{eqAx5}
\end{eqnarray}
then, inserting Eq.~(\ref{eqAx1}),
\begin{eqnarray}
 g_{\mathcal{P}\mathcal{T}}(x+1,y)&=&\eta_{xy}^y\eta_{x\mathcal{T}}\eta_{\mathcal{T}}^{-1}g_{\mathcal{P}\mathcal{T}}(y,x),\nonumber\\
 g_{\mathcal{P}\mathcal{T}}(x,y+1)&=&\eta_{xy}^x\eta_{y\mathcal{T}}\eta_{\mathcal{T}}^{-1}g_{\mathcal{P}\mathcal{T}}(y,x).
  \label{eqAx6}
\end{eqnarray}
The solution is
\begin{equation}
  g_{\mathcal{P}\mathcal{T}}(x,y)=\eta_{x\mathcal{P}\mathcal{T}}^x\eta_{y\mathcal{P}\mathcal{T}}^y\eta_{xy}^{xy}\eta_{\mathcal{P}\mathcal{T}}^{-x-y}g_{\mathcal{P}\mathcal{T}}.
  \label{eqAx7}
\end{equation}
Considering Eq.~(\ref{eqAx1}) on the site $(0,0)$, we have
\begin{eqnarray}
 g_{\mathcal{P}\mathcal{T}}(b)\mathcal{K}g_{\mathcal{P}\mathcal{T}}(a)\mathcal{K}&=&\eta_{\mathcal{P}\mathcal{T}},\nonumber\\
 g_{\mathcal{P}\mathcal{T}}(a)\mathcal{K}g_{\mathcal{P}\mathcal{T}}(b)\mathcal{K}&=&\eta_{\mathcal{P}\mathcal{T}}.
  \label{eqAx8}
\end{eqnarray}
Since $\eta_{\mathcal{P}\mathcal{T}}\in{\rm IGG}$, the equations hold only if $\eta_{\mathcal{P}\mathcal{T}}=1$. Thus, $g_{\mathcal{P}\mathcal{T}}(b)g_{\mathcal{P}\mathcal{T}}^*(a)=1$.

Substituting Eq.~(\ref{eqAx7}) into Eq.~(\ref{eqAx1}),
\begin{equation}
  \eta_{x\mathcal{P}\mathcal{T}}^{y-x}\eta_{y\mathcal{P}\mathcal{T}}^{x-y}=1.
  \label{eqAx9}
\end{equation}
Therefore, $\eta_{x\mathcal{P}\mathcal{T}}=\eta_{y\mathcal{P}\mathcal{T}}$. Then substituting Eq.~(\ref{eqAx7}) into Eq.~(\ref{eqAx4}),
\begin{eqnarray}
  \eta_{x\mathcal{P}\mathcal{T}}^{2y-x}g_{C_6}(a)g_{\mathcal{P}\mathcal{T}}(b)g_{C_6}^*(a)g_{\mathcal{P}\mathcal{T}}^*(b)&=&\eta_{C_6\mathcal{P}\mathcal{T}},\nonumber\\
  \eta_{x\mathcal{P}\mathcal{T}}^{2y-x-1}g_{C_6}(b)g_{\mathcal{P}\mathcal{T}}(a)g_{C_6}^*(b)g_{\mathcal{P}\mathcal{T}}^*(a)&=&\eta_{C_6\mathcal{P}\mathcal{T}}.
  \label{eqAx10}
\end{eqnarray}
The right-hand side of the equations are independent of $(x,y)$, so $\eta_{x\mathcal{P}\mathcal{T}}=1$. Similarly, we have $\eta_{\mathcal{P}\mathcal{T}}=1$ and $\eta_{C_6\mathcal{P}\mathcal{T}}=1$. If a proper global gauge was chosen, one can show that $g_{\mathcal{P}\mathcal{T}}(a)=g_{\mathcal{P}\mathcal{T}}(b)=1$.

In sum, there is no extra type of PSG when $\mathcal{P}\mathcal{T}$ presents.

\subsection{$\mathcal{T}$ and $\mathcal{P}$}
If $\mathcal{T}$ presents, substituting Eq.~(\ref{eqA14}) into Eq.~(\ref{eqA2}) and (\ref{eqA3}), we obtian
\begin{eqnarray}
 g_\mathcal{T}(x+1,y)\mathcal{K}g_\mathcal{T}(x,y)\mathcal{K}&=&\eta_{xy}^{-y}\eta_{x\mathcal{T}},\nonumber\\
 g_\mathcal{T}(x,y+1)\mathcal{K}g_\mathcal{T}(x,y)\mathcal{K}&=&\eta_{y\mathcal{T}},
  \label{eqA19}
\end{eqnarray}
then, inserting Eq.~(\ref{eqA8}),
\begin{eqnarray}
 g_\mathcal{T}(x+1,y)&=&\eta_{xy}^{-y}\eta_{x\mathcal{T}}\eta_{\mathcal{T}}^{-1}g_\mathcal{T}(x,y),\nonumber\\
 g_\mathcal{T}(x,y+1)&=&\eta_{y\mathcal{T}}\eta_{\mathcal{T}}^{-1}g_\mathcal{T}(x,y).
  \label{eqA20}
\end{eqnarray}
The solution exists only if $\eta_{xy}=1$, and it is
\begin{equation}
  g_\mathcal{T}(x,y)=\eta_{x\mathcal{T}}^x\eta_{y\mathcal{T}}^y\eta_{\mathcal{T}}^{-x-y}g_{\mathcal{T}}.
  \label{eqA21}
\end{equation}

As we know that for a time-reversal transformation, its square, such as Eq.~(\ref{eqA8}), is $1$ or $-1$. Since $-1$ is not a SU(3) element, only $\eta_{\mathcal{T}}=1$ holds. Choosing a proper global gauge, we can always choose $g_{\mathcal{T}}(a)=1$.

Substituting Eqs.~(\ref{eqA17}) and (\ref{eqA21}) into Eq.~(\ref{eqA11}), we have
\begin{eqnarray}
  \eta_{x\mathcal{T}}^{-y-1}\eta_{y\mathcal{T}}^{x-y}\mathcal{K}g_\mathcal{T}(b)\mathcal{K} &=& \eta_{\mathcal{T}C_6}g_{C_6}(b)^2, \nonumber\\
  \eta_{x\mathcal{T}}^{-y}\eta_{y\mathcal{T}}^{x-y}g_\mathcal{T}(b)&=& \eta_{\mathcal{T}C_6}.
  \label{eqA22}
\end{eqnarray}
The solution is
\begin{equation}
  g_\mathcal{T}(b)= \eta_{\mathcal{T}C_6},~g_{C_6}(b)= \eta_{\mathcal{T}C_6}^{-1},
  \label{eqA23}
\end{equation}
with $\eta_{x\mathcal{T}}=\eta_{y\mathcal{T}}=1$. Now there is only one independent $\eta_{ab}$: $\eta_{\mathcal{T}C_6}$, so there are three types and each type has only one solution.

Before going on, we try to figure out the solutions if there is no time-reversal symmetry. Substituting Eq.~(\ref{eqA14}) into Eq.~(\ref{eqA6}) and (\ref{eqA7}), then a similar process as for $g_\mathcal{T}$ will be given, which would lead to $\eta_{xy}=1$ and
\begin{equation}
  g_\mathcal{P}(x,y)=\eta_{\mathcal{P} x}^x\eta_{\mathcal{P} y}^yg_{\mathcal{P}}.
  \label{eqA24}
\end{equation}
Inserting it into Eq.~(\ref{eqA10}) and (\ref{eqA13}), we get
\begin{eqnarray}
  \eta_{\mathcal{P} x}^{x+y}\eta_{\mathcal{P} y}^{x+y}g_\mathcal{P}(\zeta)g_\mathcal{P}(\overline{\zeta}) &=& \eta_\mathcal{P}, \\
  \eta_{\mathcal{P} x}^{-y}\left[g_{C_6}(b)g_\mathcal{P}(a)\right]^2 &=& \eta_{C_6\mathcal{P}},\\
   \eta_{\mathcal{P} x}^{-y-1}\left[g_{C_6}(a)g_\mathcal{P}(b)\right]^2&=& \eta_{C_6\mathcal{P}}.
\end{eqnarray}
The right-hand side of these equations are independent of $(x,y)$, so must be the left-hand side. Thus $\eta_{\mathcal{P} x}=\eta_{\mathcal{P} y}=1$. After simplifying with Eq.~(\ref{eqA18}), we obtain the equations
\begin{eqnarray}
  g_\mathcal{P}(a)g_{C_6}(b)g_\mathcal{P}(a)&=& \eta_{C_6}^2\eta_{C_6\mathcal{P}}g_{C_6}(b)^2 \nonumber \\
  g_\mathcal{P}(\zeta)g_\mathcal{P}(\overline{\zeta}) &=& \eta_\mathcal{P} \nonumber \\
  \left[g_{C_6}(b)g_\mathcal{P}(a)\right]^2 &=& g_\mathcal{P}(b)^2=\eta_{C_6\mathcal{P}}.
\end{eqnarray}
The solutions are
\begin{eqnarray}
   \eta_{C_6}= 1,&&~g_{C_6}(b)=\eta_\mathcal{P}^2\eta_{C_6\mathcal{P}}. \nonumber\\
  g_\mathcal{P}(b)^2 = \eta_{C_6\mathcal{P}},&&~g_\mathcal{P}(a)= \eta_\mathcal{P}\eta_{C_6\mathcal{P}}^2g_\mathcal{P}(b).
  \label{eqA29}
\end{eqnarray}
Careful calculations will give $9$ types of PSG and their corresponding $36$ choices.

Taking both $\mathcal{P}$ and $\mathcal{T}$ into account, Eq.~(\ref{eqA11}) and (\ref{eqA12}) give
\begin{equation}
  \eta_{\mathcal{T}\mathcal{P}}\eta_\mathcal{P}=1,~\eta_{\mathcal{T}C_6}\eta_\mathcal{P}^2\eta_{C_6\mathcal{P}}=1.
\end{equation}
Thus, only two of $\eta_{\mathcal{T}\mathcal{P}},\eta_{\mathcal{T}C_6},\eta_\mathcal{P}$ and $\eta_{C_6\mathcal{P}}$ are independent, which also results in $9$ types of PSG and their corresponding $36$ choices.

\bibliography{parafermion}

\end{document}